\documentclass[conference]{IEEEtran}
\usepackage{cite}
\usepackage{algorithmic}
\usepackage{graphicx}
\usepackage{textcomp}
\usepackage{xcolor}
\usepackage{subcaption}
\usepackage[utf8]{inputenc}
\usepackage[cmex10]{amsmath}
\usepackage{amssymb}

\DeclareMathOperator*{\minimize}{minimize}
\DeclareMathOperator*{\subjectto}{subject\,to}
\usepackage{hyperref}

\hypersetup{
    colorlinks=false,
    hidelinks
    }
\usepackage{amsthm}

\usepackage{comment}

\def\BibTeX{{\rm B\kern-.05em{\sc i\kern-.025em b}\kern-.08em
    T\kern-.1667em\lower.7ex\hbox{E}\kern-.125emX}}
\begin{document}

\title{Inference Load-Aware Orchestration for Hierarchical Federated Learning}

 \author{
    \IEEEauthorblockN{ 
    Anna Lackinger\IEEEauthorrefmark{1},
    Pantelis A. Frangoudis\IEEEauthorrefmark{1}, 
    Ivan Čilić\IEEEauthorrefmark{2},
    Alireza Furutanpey\IEEEauthorrefmark{1},
    Ilir Murturi\IEEEauthorrefmark{1}, \\
    Ivana Podnar Žarko\IEEEauthorrefmark{2},
    Schahram Dustdar\IEEEauthorrefmark{1}
    }    
    \vspace{2mm}
    \IEEEauthorblockA{\IEEEauthorrefmark{1}Distributed Systems Group, TU Wien, Vienna, Austria
    \\\{a.lackinger, p.frangoudis, a.furutanpey, imurturi, dustdar\}@dsg.tuwien.ac.at} 
    \vspace{1mm}
    \IEEEauthorblockA{\IEEEauthorrefmark{2}Faculty of Electrical Engineering and Computing, University of Zagreb, Zagreb, Croatia
    \\\{ivan.cilic, ivana.podnar\}@fer.hr}
    \vspace{-1cm}
}

\maketitle

\begin{abstract}
Hierarchical federated learning (HFL) designs introduce intermediate aggregator nodes between clients and the global federated learning server in order to reduce communication costs and distribute server load. One side effect is that machine learning model replication at scale comes ``for free'' as part of the HFL process: model replicas are hosted at the client end, intermediate nodes, and the global server level and are readily available for serving inference requests. This creates opportunities for efficient model serving but simultaneously couples the training and serving processes and calls for their joint orchestration. This is particularly important for \emph{continual learning}, where serving a model while (re)training it periodically, upon specific triggers, or continuously, takes place over shared infrastructure spanning the computing continuum. Consequently, training and inference workloads can interfere with detrimental effects on performance. To address this issue, we propose an inference load-aware HFL orchestration scheme, which makes informed decisions on HFL configuration, considering knowledge about inference workloads and the respective processing capacity. Applying our scheme to a continual learning use case in the transportation domain, we demonstrate that by optimizing aggregator node placement and device-aggregator association, significant inference latency savings can be achieved while communication costs are drastically reduced compared to flat centralized federated learning.
\end{abstract}

\begin{IEEEkeywords}
Federated learning, service orchestration, continual learning, edge computing.
\end{IEEEkeywords}

\section{Introduction}

The traditional approach to machine/deep learning (ML/DL) is to train an ML model with data collected on a centralized cloud server and make it available for inference either by hosting it in the cloud or, more recently, serving it directly from end devices following the edge computing paradigm~\cite{Deng20}. Both training and inference are, however, challenged in different ways. 
On the one hand, data often tend to be distributed across various devices, and collecting all data to a central server raises logistical and privacy concerns~\cite{Lim2023Decentralized}. 
On the other hand, and despite the proliferation of AI accelerators of different scales, serving inference requests at the same time in a timely, accurate, and private manner is a hard, multi-faceted problem that calls for sophisticated edge-cloud synergy~\cite{frankensplit}. 

Federated learning (FL) has emerged as a response to some of these challenges and has gained significant popularity. 
In FL, clients train an ML model on local data and upload the result of their computation to a central server for aggregation. 
The server then produces and redistributes an updated global model, starting a new training round.
This iterative process continues until some convergence criterion is met. FL operates under distinctly different assumptions than traditional deep learning: the number of participating nodes is often high, data are unbalanced and often not independent and identically distributed (non-IID), and performance depends on the communication capabilities of participating entities, which are often heterogeneous, resource-constrained and unreliable~\cite{9763486}. Moreover, even when the number of participating devices is high, each device typically only holds a relatively small data volume. 

In its default centralized configuration, FL can suffer from increased network costs. Instead of uploading raw data, clients must exchange ML model copies with the FL server, which is particularly costly for long-lasting training processes and large DL models. At the same time, large-scale synchronous federated learning tasks require significant network capacity at the server end to deal with simultaneous model updates~\cite{Wong23}.
These limitations motivate \emph{hierarchical} federated learning (HFL), where frequent aggregations can be enabled on the edge network, avoiding extensive cloud aggregations without sacrificing learning accuracy~\cite{liu_hiear,Heydar20,Wang21,Xu23,Wu23,Guo23,Yinghui24,Pervej24,Trindade24,Deng24}. 

In a hierarchical FL process, the key decision is how to appropriately cluster FL clients in groups and designate local aggregator nodes per cluster to optimize for specific performance criteria, such as communication cost minimization or FL convergence time. As a side effect, ML model replication at hosts in the computing continuum comes ``for free'' as a natural part of the HFL process: besides client devices and the cloud, model replicas are now found at local aggregation points and are usable for serving inference requests, potentially with low latency due to proximity, and provided that privacy requirements allow it. At the same time, though, this creates a coupling between training and inference and calls for their \emph{joint orchestration}. This coupling becomes more important in \emph{continual learning (CL)} scenarios~\cite{shenaj2023asynchronousCL, CRIADO2022263}, where training and inference processes overlap and compete for compute and network resources. The CL paradigm provides a realistic setting reflecting real-world learning processes, with new tasks and new data, potentially with different distributions, emerging over time. As such, CL may involve retraining an ML model while serving it, which is challenging to achieve with high performance when sharing the same resource-constrained infrastructure, as is often the case in FL. This challenge has started receiving attention only recently~\cite{han2023federated}.

Our work focuses on continual hierarchical FL and contributes to answering the following: \emph{
How can a hierarchical FL process be orchestrated with inference workload awareness? Additionally, what are the performance benefits of this approach in terms of reducing inference latency and communication costs?}\\
In summary, our main contributions are:
\begin{itemize}
    \item Introducing an architectural framework for HFL orchestration~(\S\,\ref{sec:architecture}), in the context of which we formulate and solve the \emph{inference-aware HFL Orchestration Problem (HFLOP)}. This leads to a cost-optimal assignment of FL devices to aggregator nodes, taking into account their inference workload processing capacities~(\S\,\ref{sec:hflop}). To the best of our knowledge, this is the first work that tackles joint training-inference orchestration in the \emph{hierarchical} FL setting.
    \item Applying and evaluating our scheme to a continual federated learning use case in the transportation domain using a real-world dataset to handle dynamic and evolving traffic patterns~(\S\,\ref{sec:evaluation}).
    Our results show that our approach can feasibly reduce inference latency and communication costs. 
    \item Providing our code as open source\footnote{\url{https://github.com/pfrag/hflop}}  for reproducibility and to allow other researchers to utilize it for their purposes.
\end{itemize} 

\section{Related Work}
\label{sec:rwork}

\subsection{Continuous federated learning}
Continuous learning, also known as lifelong learning, has been proposed by Thrun \cite{thrun1995lifelong} and is a well-researched topic with extensive literature. It involves learning continuously from streaming data, developing a model based on previous learning, and then being able to reapply, adapt, and generalize it to new situations. Continuous learning and continual learning are often overlooked in federated learning, which is typically limited to one-time training. In practice, however, a model may need to be retrained over time to maintain its accuracy. Wang et al. \cite{wang2024comprehensive} present a comprehensive survey on continual learning, mainly focusing on the theoretical foundations, representative methods, and practical applications to address catastrophic forgetting \cite{mccloskey1989catastrophic,li2017learning, kirkpatrick2017overcoming} and maintain model performance. Le et al. \cite{le2021federated} present a federated continuous learning scheme (FCL) based on broad learning (FCL-BL), which proposes a weighted processing strategy to solve the catastrophic forgetting problem and develops a local-independent training solution for fast and accurate training. For an overview of existing FCL approaches, related challenges, and classification tasks, the reader is referred to the work of Yang et al.~\cite{Yang24}. 
We should note that the aforementioned studies \emph{do not focus on service orchestration aspects of FCL processes and do not address resource contention between (continuous) training and inference}.

\subsection{Joint training-inference optimization in federated learning}
Most research on federated learning focuses on model training (e.g., model pruning \cite{wen2022federated}, gradient compression \cite{jiang2023heterogeneity}, etc.) rather than the inference stage (e.g., \cite{zhang2023intelligence}). Especially in continuous learning, where models need to be constantly updated, it can happen that inference requests and training overlap.
One of the first papers that emphasizes the problem of joint optimization of model training and inference is by Han et al.~\cite{han2023federated}.
To characterize the clients' inference performance, the authors introduce the notion of the age of the model to represent how much the models on the client side differ from the global model.
Another challenge mentioned in the paper is the close coupling between clients' decisions, including participation probability in FL, model download probability, and service rates.
To address these challenges, the authors offer an online problem approximation that reduces complexity and optimizes resources for model training and inference. \emph{Our work shares a similar motivation but focuses on the hierarchical setting}. Despite the extensive literature on HFL~\cite{liu_hiear,Heydar20,Wang21,Xu23,Wu23,Guo23,Yinghui24,Pervej24,Trindade24,Deng24}, we are not aware of any works that jointly consider training and inference serving in HFL.

\subsection{Applications to traffic flow prediction}
From a use-case perspective, one area where continuous learning and federated learning are often required is Traffic Flow Prediction (TFP), which is also the application scenario we experiment with in this work~(\S\,\ref{sec:evaluation}). Predicting traffic flow is crucial for reducing congestion and improving transportation efficiency in smart cities~\cite{yuan2022fedstn}. In TFP, distributed sensors measure traffic flow and train local models to predict how the traffic flow will evolve in the future. Due to the constantly changing traffic situation, the models have to be retrained continuously. 
Zhang et al.~\cite{zhang2021communication} present CTFed, a hierarchical FL approach to train Graph Neural Networks for TFP, clustering clients based on the similarity of their local models. 
Liu et al.~\cite{liu2020privacy} propose an FL-based gated recurrent unit neural network algorithm (FedGRU). Yuan et al.~\cite{yuan2022fedstn}, on the other hand, propose an FL scheme based on the Spatial-Temporal Long and Short-Term Networks (FedSTN) algorithm for TFP. 
To integrate external factors into TFP, a Semantic Capture Network (SCN) is introduced to capture the regional flow relationship and the influence of Point of Interest (POI) on TFP. Hu et al.~\cite{hu2023federated} propose an FL-based spatial-temporal traffic demand prediction model (FedSTTDP) while ensuring data security and privacy. In contrast to the mentioned works, %
\emph{we also target low-latency TFP serving for the hierarchical FL setup in particular.}

\section{System architecture}
\label{sec:architecture}

Hierarchical federated learning aims to reduce centralized FL's communication cost and alleviate the load of a single aggregation server. FL clients are organized in clusters based on criteria such as their network proximity and perform aggregations first on a local level with the mediation of a local aggregation server: client devices transmit their trained models to local servers after a set number of training epochs and receive updated (aggregated) models by their local servers; this process constitutes a \emph{local aggregation round}. After a configurable number of epochs and local aggregation rounds, the derived cluster-local model is sent to the global server for global aggregation and redistribution of a new version of the global model to clusters; this constitutes a \emph{global aggregation round}.

\begin{figure}[htbp]
\centerline{\includegraphics[width=\linewidth]{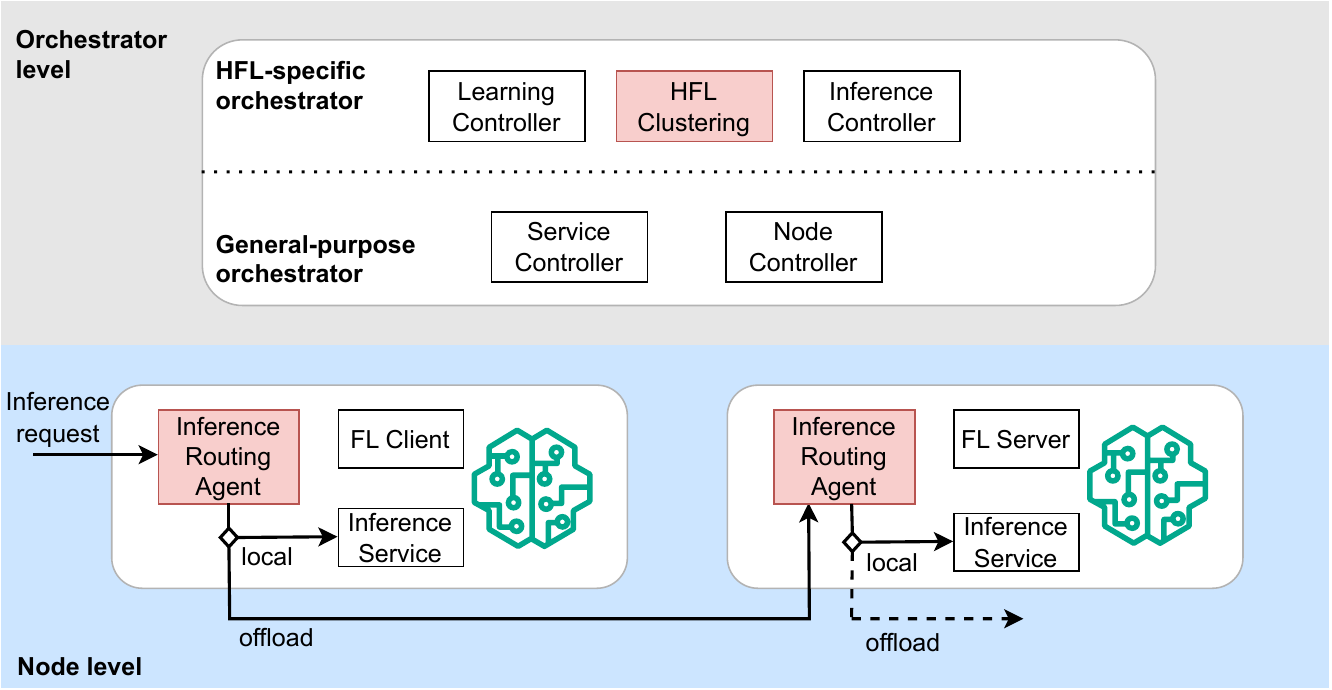}}
\caption{Architectural view of our hierarchical federated learning orchestration framework. Our framework supports the joint orchestration of distributed training and inference serving.}
\label{fig:orch}
\end{figure}

We address the challenges associated with the dynamic orchestration of such processes and put our work in the context of an HFL orchestration system. While the technical details of such a system are outside the scope of this paper, in this section, we provide its high-level overview and demonstrate how the functionality we present fits into this framework. We apply a two-level orchestration approach, depicted in \figurename~\ref{fig:orch}. Our design relies on two main components. The first is a \textit{general purpose orchestrator} (GPO), such as Kubernetes, to enable the collection of information about the underlying compute infrastructure, e.g., nodes capable of hosting HFL service components, and manage the deployment of these components as containerized microservices distributed over the computing continuum. The second is a special-purpose, \textit{HFL-specific service orchestrator}. This component makes decisions at the HFL level, and the mechanisms we present in this paper are within its scope. 

One element of this service-specific orchestrator is the \emph{learning controller}, which manages the HFL runtime. This controller retrieves information from the GPO on node resource states, underlying networks, and information on the clients requiring inference services. This information is provided as the input for the \emph{HFL clustering} mechanism, whose role is to form a hierarchy of FL clients and local aggregation nodes, solving the optimization problems therein -- a major focus of this paper is on this mechanism in particular (see Section~\ref{sec:hflop}). The output of the clustering mechanism prescribes a specific HFL configuration, which is eventually translated to (i) an actual containerized deployment implemented by the GPO at the instruction of the learning controller and, in turn, (ii) a concrete distributed HFL service instance, managed by the learning controller and implemented by microservices running on client devices, edge nodes, and the cloud. Besides assigning FL devices to aggregator nodes, and depending on the specific instantiation of our orchestration system, the clustering mechanism may decide on further FL service parameters, such as the number of local epochs and local/global aggregation frequencies. 

To enable model serving, the \emph{inference controller} is triggered at the end of the deployment phase to deploy and configure an inference service and an inference request agent per node.
In the context of hierarchical FL, an FL service can be a client that trains on local data or a server (aggregator) that provides model aggregation on a local or global level. The same model, updated during the learning process, can be used for inference. Therefore, on the same node, the orchestrator deploys an inference service, which is a server that provides an interface for interaction with the model. Also, each node hosts an inference routing agent, a proxy that decides where an inference request will be processed, implementing specific request routing logic. For example, if the node is under a heavy workload, such as when training a DL model, the routing agent forwards the requests to an inference service higher up in the hierarchy, i.e., the parent server (aggregator). Information on the load can be extracted from the GPO node controller.

We should finally note that orchestration tasks involve deploying and managing services at runtime. Therefore, orchestration must adapt to environmental changes, such as node failures and network changes. The learning controller monitors the learning pipeline and reacts to these changes by performing re-clustering on environmental events. On the other hand, orchestration also needs to adapt to service-specific events. For example, in the continual learning settings our architecture targets, a task of the inference controller is to monitor inference services and trigger a new HFL task if inference accuracy is below a specific threshold.

\section{The inference-aware HFL orchestration problem}
\label{sec:hflop}
We now focus on the internals of the HFL clustering component of our architecture. We introduce a model of our system that captures the co-existing and inter-dependent training and inference processes and we provide an integer linear programming formulation for the inference-aware HFL orchestration problem. A solution to this problem represents a cost-minimizing HFL cluster configuration.
\subsection{System model}
A set of $n$ devices participate in an FL task, and $m$ edge host locations exist that are eligible to place an aggregator.
The communication cost for device-edge host pair $(i,j)$ is given by $c^{(d)}_{ij}$. For example, $c^{(d)}_{ij} = 0$ implies that uploading data from device $i$ to an aggregator placed at edge node $j$ incurs no cost, as would be the case for an aggregator placed inside a device's local area network or, more generally, if the device is connected over an unmetered link. Similarly, the communication cost between the global server and edge node $j$ is given by $c^{(e)}_j$.

After several \emph{epochs}, clients upload the result of their computation to their associated aggregator (\emph{local round}). After a specified number of \emph{local} rounds, all aggregators submit their models to the global server, which computes a global model and broadcasts it to the aggregators, which in turn relay it to their associated client devices (\emph{global round}). 

At the same, device $i$ generates inference requests at a rate of $\lambda_{i}$. Each inference request must be processed by a node hosting a version of the ML model under training, i.e., the device itself, its associated aggregator, or the global server. Edge node $j$ has a specific inference request processing capacity $r_j$ (in requests/s), while we assume for simplicity that the capacity of the cloud-hosted global server is infinite. Requests are routed using the following rules:
\begin{itemize}
    \item \textbf{R1:} If the device generating an inference request is busy training, the request is always offloaded to the device's associated aggregator.
    \item \textbf{R2:} If the device is not participating in the current FL round, it independently decides to process the request locally or offload it to the closest aggregator.
    \item \textbf{R3:} The aggregator processes requests of its associated busy devices with priority and may serve requests by devices not active in the current FL round (or, more generally, requests by entities external to the FL process) only if the load generated by the former is sufficiently below its capacity. Otherwise, it will offload excess requests to the cloud; the aggregator thus operates as a device \emph{proxy}.
\end{itemize}

\subsection{Problem formulation}
Given these inference workload processing rules, the HFL orchestrator needs to solve the following problem: \emph{How to place aggregators at edge nodes and assign FL clients to them so that FL communication cost is minimized, subject to inference processing capacity constraints of edge hosts.} 

A solution to this problem is represented by assigning values to binary variables $x_{ij}$ for $i = 1, 2, \cdots, n$ and $j = 1, 2, \cdots, m$. These variables encode device-edge node associations, where $x_{ij} = 1$ if device $i$ is associated with an aggregator placed at edge node $j$, and $x_{ij} = 0$ otherwise. We further introduce binary variables $y_j, j = 1, 2, \cdots, m$, where $y_j = 1$ if an aggregator is placed at edge node $j$ and 0 otherwise. 

We provide the following binary integer linear programming formulation for the \emph{HFL Orchestration Problem (HFLOP)}, where the objective is to find a configuration that minimizes the communication costs of HFL:
{\allowdisplaybreaks
\begin{align}
	\minimize & \hspace{0.5cm} \sum_{i=1}^{n} \sum_{j=1}^{m} x_{ij} c^{(d)}_{ij} l + \sum_{j=1}^{n} y_{j}c^{(e)}_{j}\label{eq:problem}\\
    \subjectto & \hspace{0.5cm} x_{ij} \leq y_j, \;\; \forall\; i \in \left[1,n\right],\; \forall\; j \in \left[1,m\right] \label{c:aggregator-indicator-1}\\
     & \hspace{0.5cm} y_j \leq \sum_{i=1}^{n} x_{ij}, \;\; \forall\; j \in \left[1,m\right] \label{c:aggregator-indicator-2}\\
    & \hspace{0.5cm} \sum_{i=1}^{n} x_{ij}\lambda_i \leq r_j, \;\;\forall \;j \in \left[1,m\right] \label{c:capacity}\\
	& \hspace{0.5cm} \sum_{j=1}^{m} x_{ij} \leq 1,\;\;  \forall \; i \in \left[1,n\right] \label{c:at-most-one}\\
 	& \hspace{0.5cm} \sum_{i=1}^{n}\sum_{j=1}^{m} x_{ij} \geq T \label{c:minimum-participation}\\
	& \hspace{0.5cm} x_{ij}, y_j \in \{0,1\},\; \forall \;i\in[1,n], \; \forall \;j\in[1,m]\label{c:binary}
\end{align}
}

The objective function~(\ref{eq:problem}) consists of two terms. The LHS term expresses the cost of local aggregation rounds, where $l$ is the number of local aggregations taking place before a global aggregation round. In practical terms, \emph{assigning devices to local edge aggregators}, to which connectivity is cheap/free, tends to reduce this cost component. The RHS term is the cost of global aggregation, which involves each edge aggregator uploading a model update to the global server and receiving the global model. \emph{Reducing the number of edge aggregators} tends to reduce this cost (limited by edge host capacity constraints and at the expense of potentially having clients associated with remote aggregators).

We have dropped the constants expressing the model size and that both local and global aggregation involve two communication directions; thus, in absolute terms, the cost doubles. Finally, we assume the uplink and downlink costs are the same, though our model is straightforward to extend in order to capture cost asymmetries.
Constraint set~(\ref{c:aggregator-indicator-1}) ensures that an aggregator is always placed at edge node $j$ if at least one edge device is associated with it. 
Conversely, constraint set~(\ref{c:aggregator-indicator-2}) does not allow an aggregator to be placed at an edge node without any associated devices. This set of constraints is often redundant, as operating an aggregator at node $j$ typically comes at a cost; if no devices are associated with it, then $y_j = 1$ will not be part of the optimal solution. 
Edge node capacity constraints are given by~(\ref{c:capacity}). 
Constraint~(\ref{c:at-most-one}) allows each device to associate with at most one aggregator. 
Additionally, constraint~(\ref{c:minimum-participation}) ensures that a minimum number of devices $T$ participates in FL. 
Finally,~(\ref{c:binary}) constrains the variables to binary values.

Notably, HFLOP is a generalization of the \emph{Capacitated Facility Location Problem (CFLP) with unsplittable flows}, i.e., the CFLP version where a single facility must serve the whole demand of an entity. This problem is known to be NP-hard~\cite{Shmoys97}. Any instance of this CFLP version can be transformed to an instance of HFLOP by (i) creating one edge node for each potential facility location and one device for each location to be served, (ii) setting $c^{(e)}_j$ to the setup cost of each corresponding location, and device communication costs $c^{(d)}_{ij}$ to the transportation costs between locations $i$ and $j$, and (iii) setting $T = n$, i.e., all devices (cf. locations) need to be covered.

\subsection{Performance considerations}
\begin{figure}
    \includegraphics[width=0.9\columnwidth]{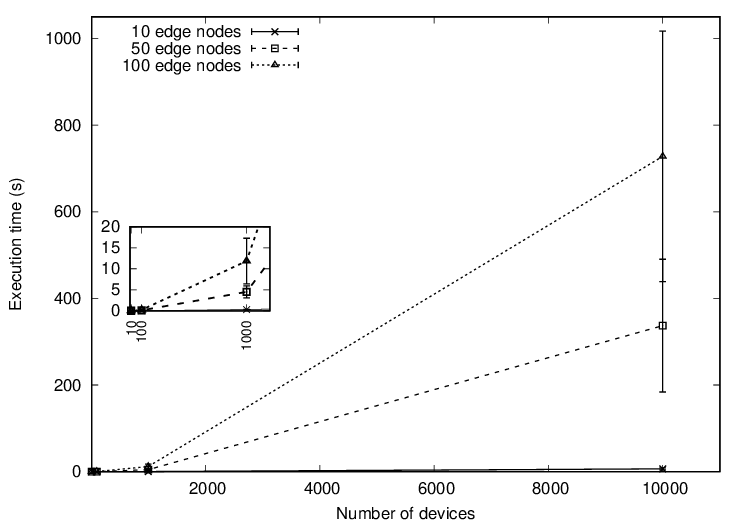}
    \caption{Execution times of deriving the optimal solution to HFLOP using a commercial solver (CPLEX). Mean execution times are reported with 95\% confidence intervals.}
    \label{fig:hflop-execution-time}
\end{figure}

Given the hardness of the problem, dealing with large instances can be challenging. \figurename~\ref{fig:hflop-execution-time} presents the time to derive the optimal solution to HFLOP for varying problem instance sizes using the branch and cut algorithm by CPLEX on a host with an 8-core AMD Ryzen CPU and 32\, GB of RAM. For scenarios with 10.000 devices and 100 edge host locations, it takes hundreds of seconds to find an optimal assignment, while for larger instances, it can become prohibitively expensive computationally to solve the problem optimally due to severely increasing CPU and memory requirements. However, this approach is feasible for many practical cases, given that HFLOP may not need to be solved frequently. For larger-scale scenarios, adaptations of heuristics and approximation algorithms for versions of the facility location problem~\cite{Shmoys97,Arya01} can be considered.

\section{Evaluation}
\label{sec:evaluation}
The following experiments aim to evaluate the performance of our scheme (HFLOP) on a real-world dataset.
We wish to quantify the effects of using HFLOP on the accuracy of federated learning clients, the communication cost of aggregation rounds, and the inference serving time.
By examining these factors, we can determine whether HFLOP improves communication costs and inference processing time without negatively impacting training. 

\subsection{Methodology and use case}

\begin{figure}[htbp]
\centerline{\includegraphics[width=\linewidth]{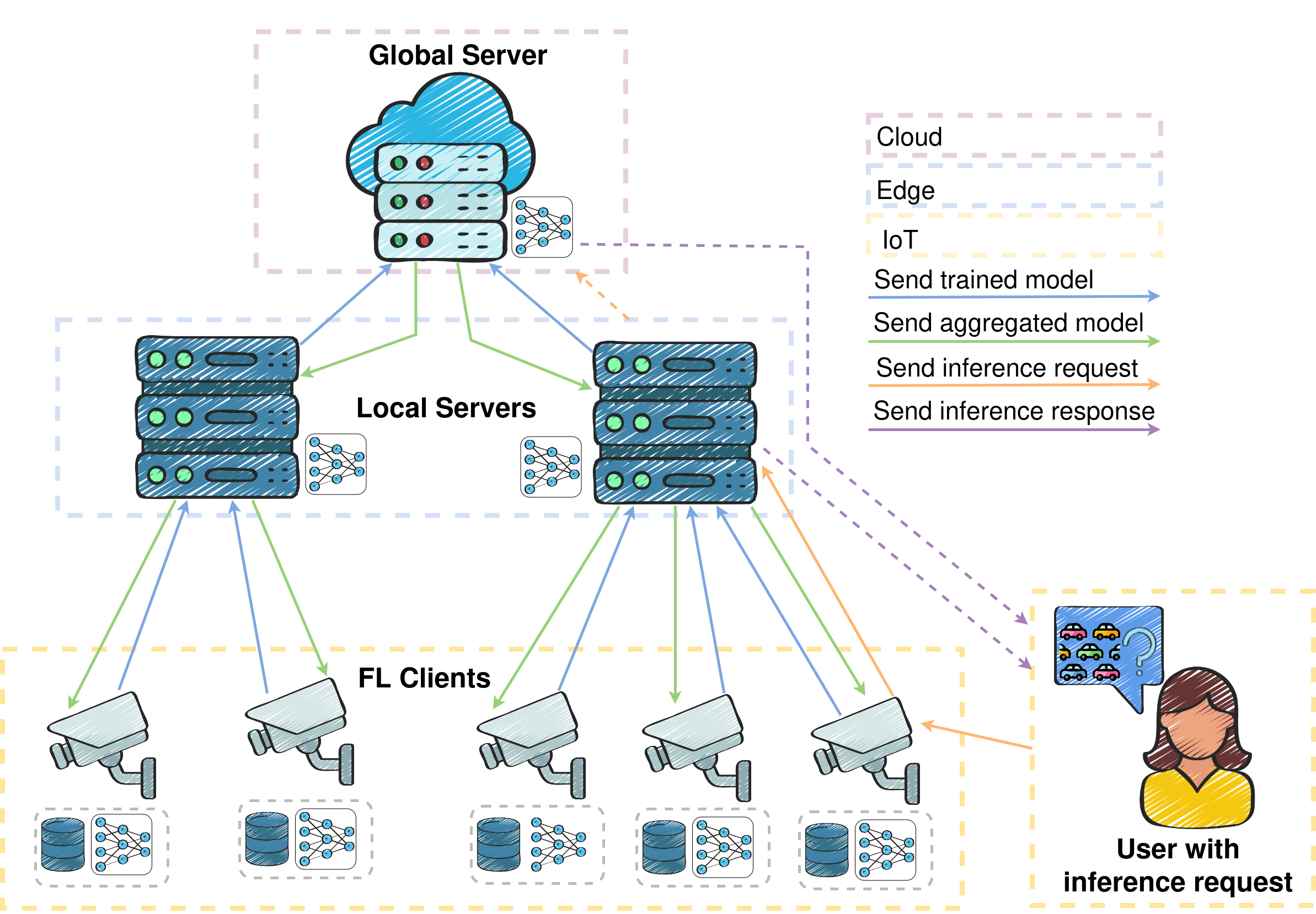}}
\caption{
Different FL clients are clustered and connected to their closest local server. 
An inference request is sent to the closest server, 
which processes the request on its model or sends it to a local server if it is busy training. 
The local server processes the request or, if it reaches its processing capacity, forwards it to a cloud server, which then answers the request.}
\label{fig:system}
\end{figure}

\figurename~\ref{fig:system} illustrates a hierarchical federated learning setup with inference requests and an overview of the use case that shapes our experiment setup.
Federated learning clients continuously train models on their local data. After some time, they send their data to local servers that either aggregate the models and send them back to the clients or forward the aggregated models to the global server.
While some federated learning clients are training, some entities may send inference requests to the closest server, which is a federated learning client in most cases.
The inference request will be forwarded to the closest local server if the FL client is busy training. 
If this server is busy, the request will be forwarded to the global server.

\begin{figure}[htbp]
\includegraphics[width=\linewidth, trim=100 150 100 130, clip]{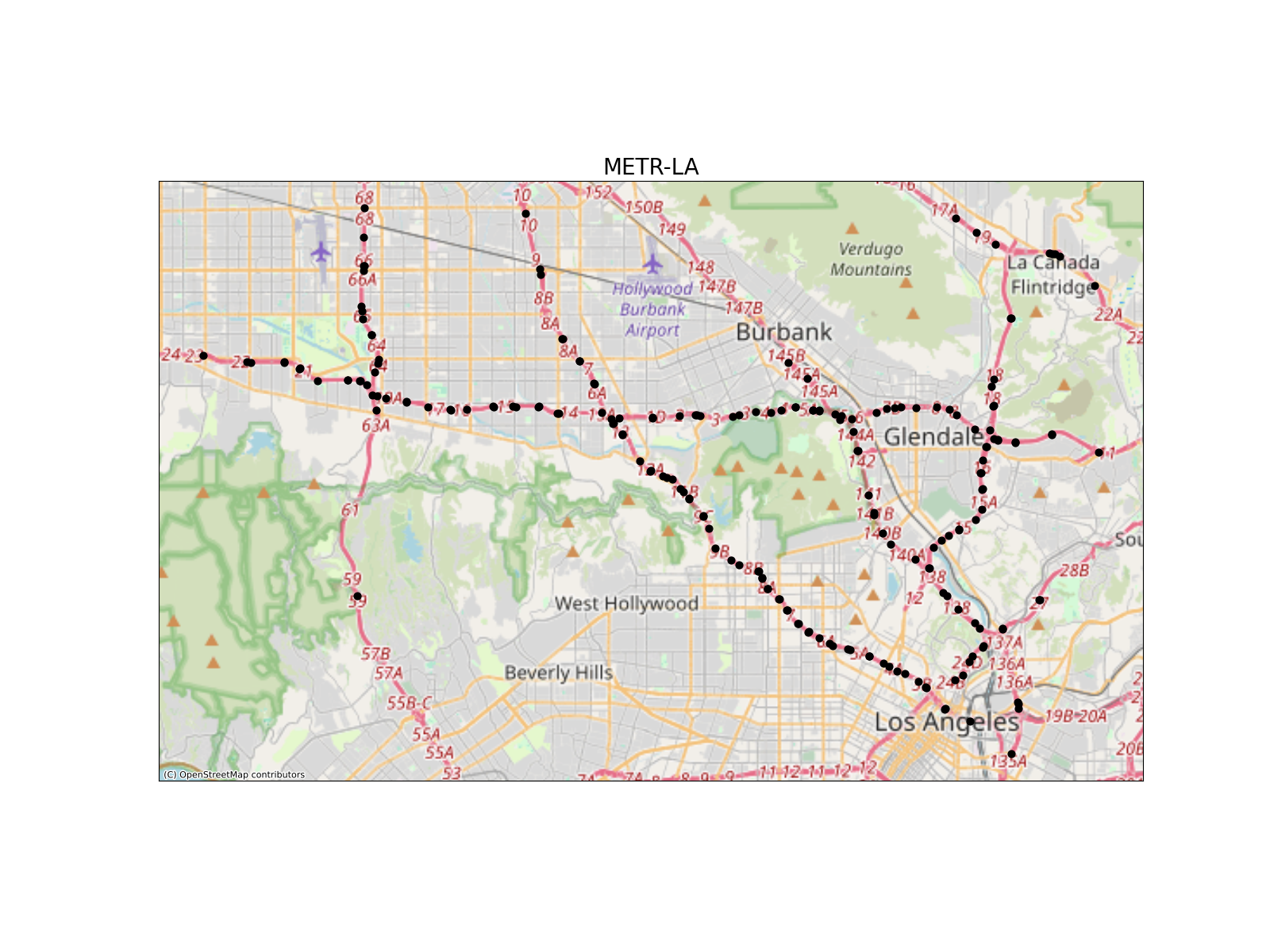}
\caption{Sensor distribution of the METR-LA dataset.}
\label{fig:streetmap}
\end{figure}

We use the traffic forecasting dataset based on Los Angeles Metropolitan traffic (METR-LA) for the experiments.
Its sensors are depicted in \figurename~\ref{fig:streetmap}. 
This dataset contains traffic information collected from loop detectors in the highway of Los Angeles County \cite{li2018diffusion}. 
In the dataset, there are 207 sensors and 4 months of data ranging from March 1st, 2012, to June 30th, 2012.
Traffic information is recorded every 5 minutes for 34.272 time stamps.
\vspace{-0.1cm}

\vspace{-0.4cm}

\subsection{Continuous learning performance}
\subsubsection{Continuous learning} 
The tests conducted in this work for continuous learning, continuous federated learning, and continuous federated learning with inference requests ran on a server that is equipped with 32 CPUs, specifically the AMD Ryzen 9 5950X 16-Core Processor, each with a maximum clock speed of 5083.3979 MHz. 
Additionally, the server incorporates 2 NVIDIA GeForce RTX 3090 GPUs, each featuring 24576 MB of memory.

To test the benefits of continuous learning and whether it improves the accuracy of the model, we conducted some experiments using a Gated Recurrent Units (GRU) architecture.
Recurrent Neural Networks (RNNs) are often used for time-series predictions, and GRUs have gating mechanisms to improve the information flow. Therefore, they are often used in federated time-series prediction \cite{li2018diffusion,liu2020privacy}.
The GRU was trained for 20 epochs on 4 weeks of the data, and the remaining data was used for testing. 
The best results were achieved using a batch size of 16, a hidden size of 128, 2 layers, and a learning rate of 0.0001.
After training for 100 epochs, the Mean Squared Error (MSE) predicting the test data was 0.04470, and the MSE of the retrained model that continuously retrained on new data was 0.04284.
These results demonstrate that continuously training the model decreased the MSE.

\subsubsection{Continuous federated learning}
\label{sec:continuous-fl}
After the first tests were conducted to find the best hyperparameters for the model and to show the benefits of continuous training, the next step was to evaluate continuous federated learning.
We first clustered the clients for the hierarchical experiments based on their location.
This ensures that when an edge aggregator is added, its network cost and latency are reduced.
In \figurename~\ref{fig:clustered}, one can see the street map after the sensors have been clustered.
Each color represents a different cluster.
The minimal number of participating devices given in Constraint~({\ref{c:minimum-participation}) is for these experiments set to 20.
Since there are 4 clusters, 5 random sensors were chosen from each cluster for training an ML model. 
These sensors that were chosen are marked in red and are also shown in \figurename~\ref{fig:clustered}.
\vspace{-0.1cm}
\begin{figure}[htbp]
\includegraphics[width=\linewidth, trim=100 150 100 130, clip]{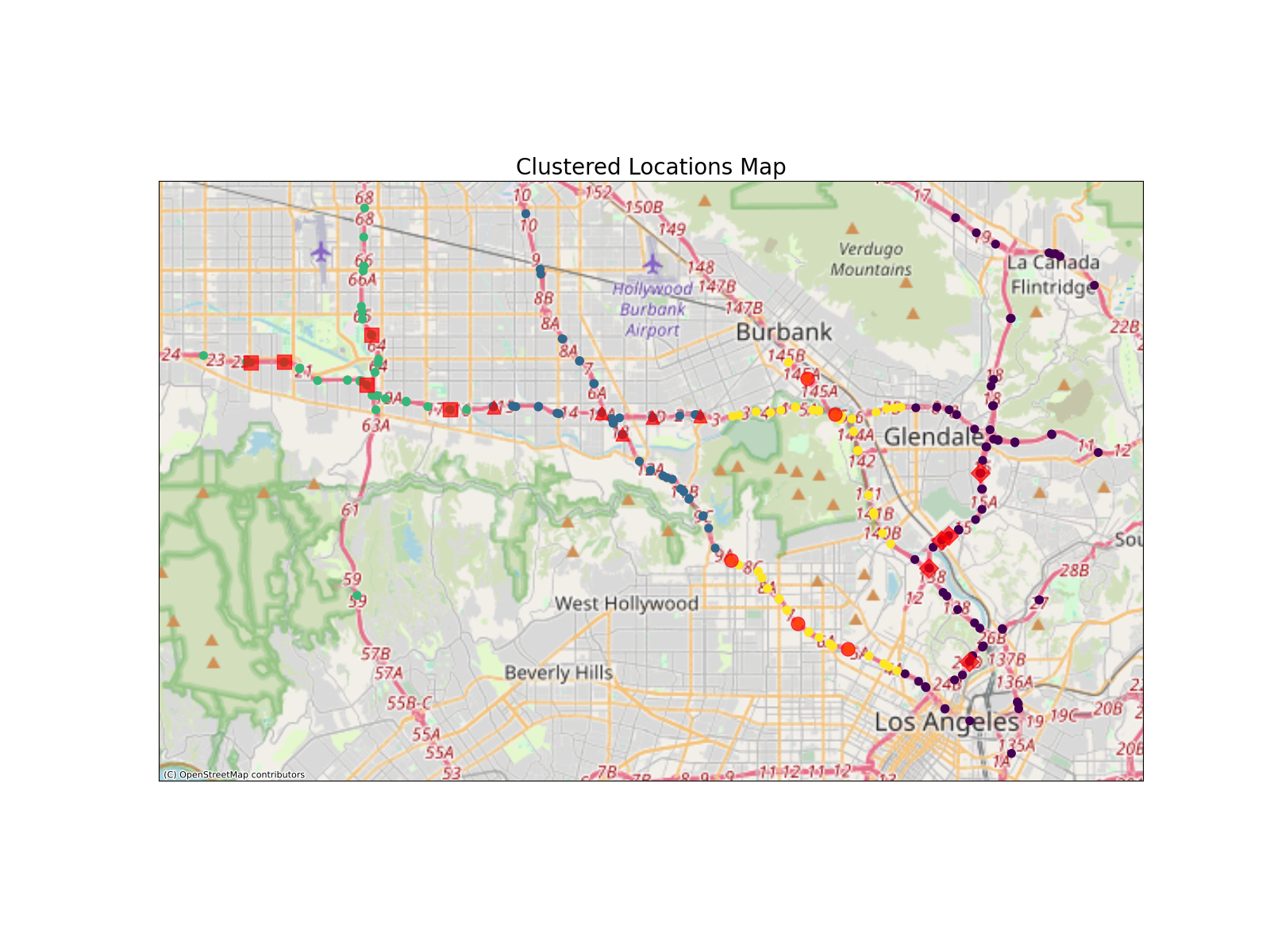}
\caption{Clustered sensors using the METR-LA dataset.}
\label{fig:clustered}
\end{figure}
\vspace{-0.1cm}

To simulate continuous learning, we use 3 weeks of training and 1 week of validation.
After each aggregation round, the global time shifts for some timestamps so that the number of training and test samples stays the same, but it is shifted to simulate time passing.
The MSE results for 100 aggregation rounds, using 20 clients trained for 5 epochs and 3 different setups, are shown in \figurename~\ref{fig:global}.

\begin{figure}[htbp]
    \centering
    \begin{subfigure}[b]{0.24\textwidth}
        \centering
        \includegraphics[width=\linewidth]{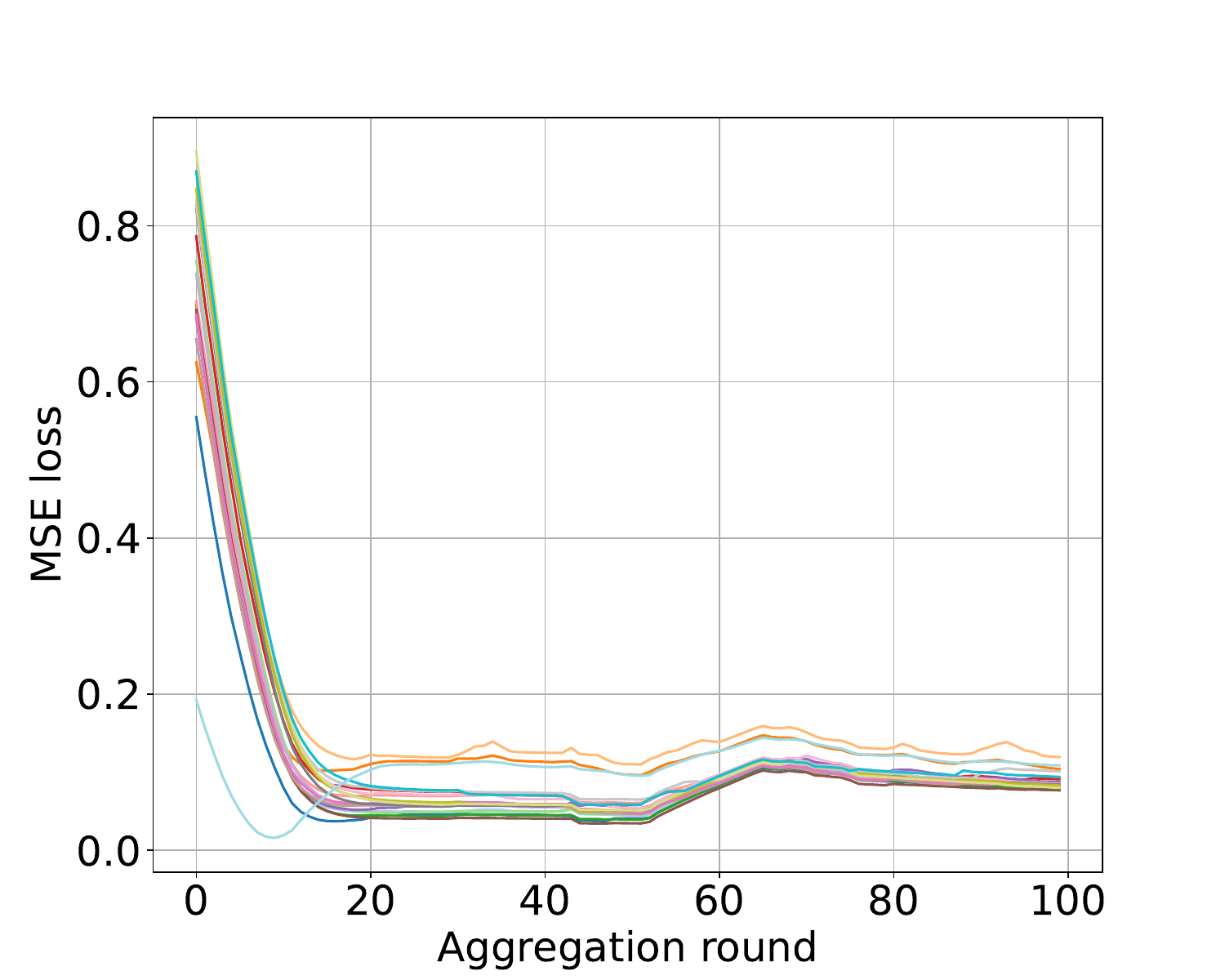}
        \caption{Non-hierarchical setup.}
        \label{fig:nonhier}
    \end{subfigure}
    \vfill
    \begin{subfigure}[b]{0.24\textwidth}
        \centering
    \includegraphics[width=\linewidth]{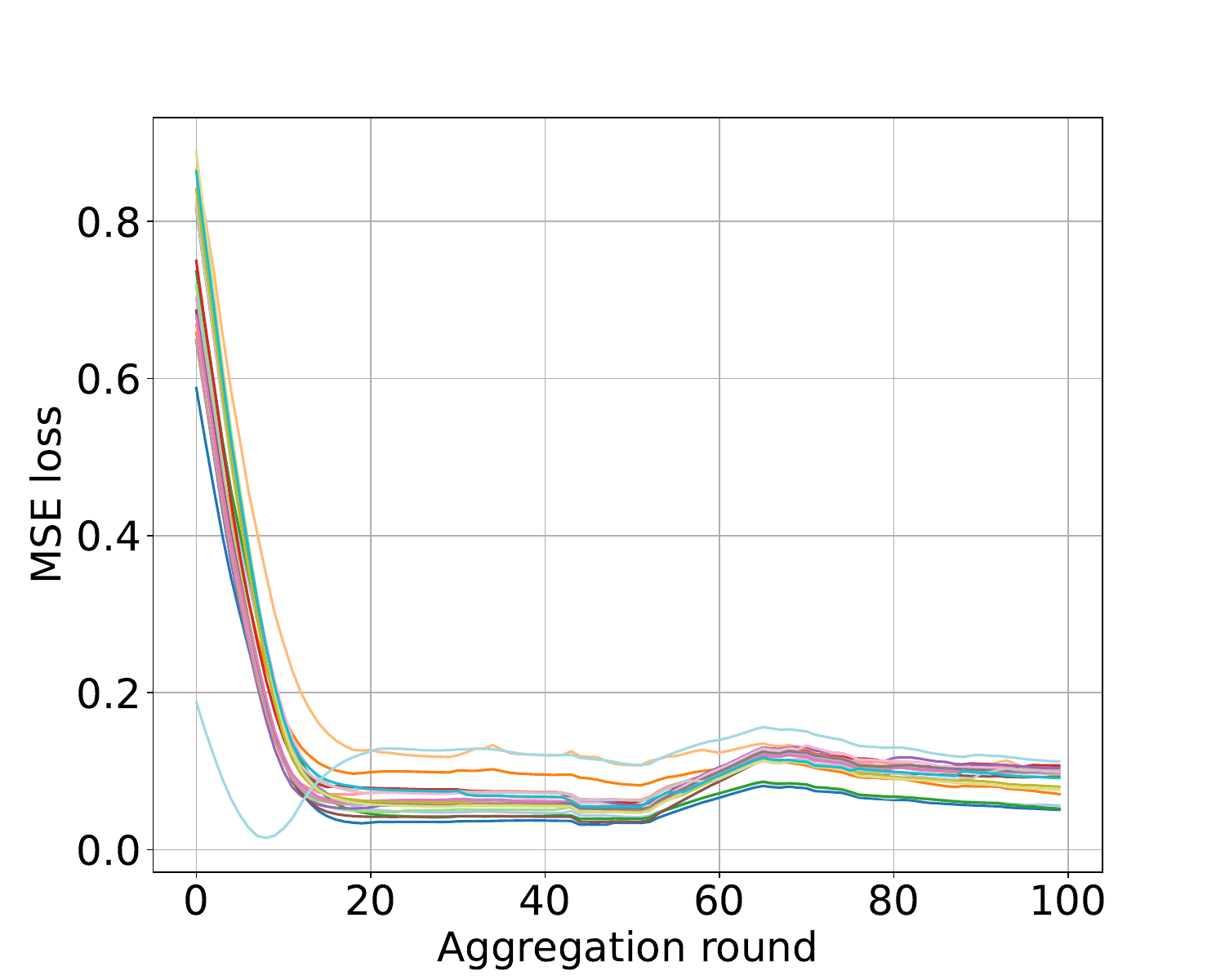}
        \caption{Hierarchical setup with 4 edge servers and 2 local aggregations before a global aggregation.}
        \label{fig:hier}
    \end{subfigure}
    \begin{subfigure}[b]{0.24\textwidth}
        \centering
        \includegraphics[width=\linewidth]{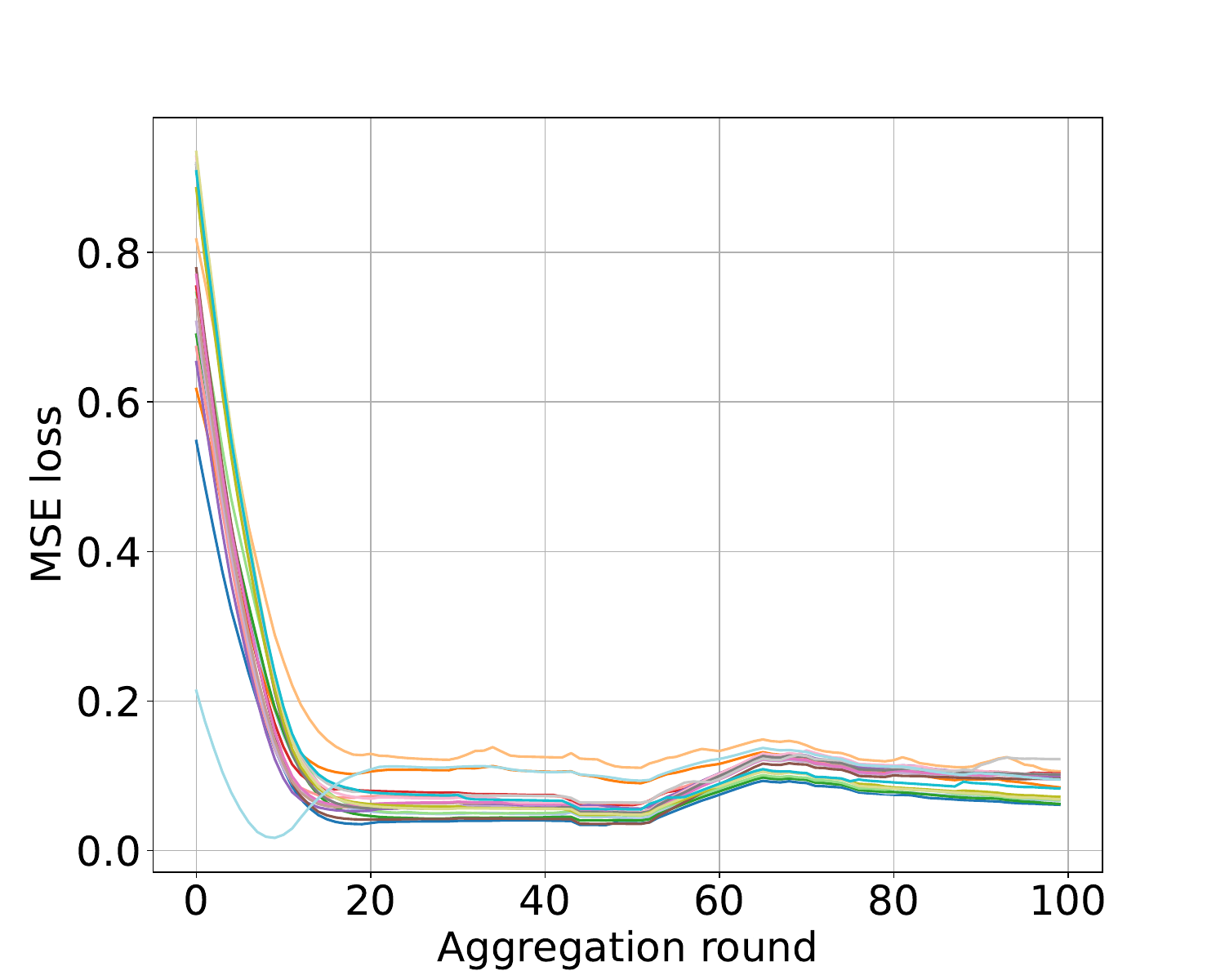}
        \caption{HFLOP setup with 4 edge servers and 2 local aggregations before a global aggregation.}
        \label{fig:hflop}
    \end{subfigure}
    \caption{MSE of 20 clients, each of them is represented in a different color. Each client tests on its local data after receiving the global model.}
    \label{fig:global}
\end{figure}

In this figure, one can see the MSE that has been calculated for each client directly after receiving the global model for up to 100 aggregation rounds when having a non-hierarchical setup (\figurename~\ref{fig:nonhier}), a hierarchical setup based only on the locations (Figure \ref{fig:hier}) and a hierarchical setup using HFLOP (\figurename~\ref{fig:hflop}).
\figurename~\ref{fig:global} demonstrates that the global MSE improves after the first few aggregation rounds as the MSE significantly decreases for most clients.
For all settings, the model converges after about 20 aggregation rounds; for the non-hierarchical setup, this means that about 20 aggregation rounds with the cloud server are needed, whereas in the hierarchical setups, 20 aggregation rounds with the local servers were required and only 10 cloud aggregations.
The MSE slightly increases after around 50 aggregation rounds.
One reason for this increase may be the changing data.
However, after some time, the MSE decreases again, showing that there are some oscillations, but overall, the model indeed converges.

\subsection{Inference serving performance}
\subsubsection{Benchmarks}
The subsequent experiments demonstrate the performance when inference requests are served while the clients continuously train.
The clustering strategies for the hierarchical experiments remain the same. 
In addition, each of the federated learning devices is assigned a rate $\lambda_{i}$ at which it produces inference requests.
For the hierarchical federated learning experiments, edge servers are added, each having a certain capacity $r_j$ to handle inference requests.
If an edge server receives more inference requests than it can handle, it forwards the additional requests to the cloud server.

\paragraph{Non-hierarchical FL benchmark}
In the non-hierarchical experiment, the federated learning clients busy with training forward their inference requests to a cloud server.

\paragraph{Hierarchical FL benchmark}
For the hierarchical FL benchmark experiments, the clients are clustered based on their location and send their inference requests to their local aggregator. 
If the number of inference requests exceeds capacity, the local aggregator forwards the request to the global aggregator.

\paragraph{Hierarchical FL with HFLOP}
For these experiments, the clients are clustered using HFLOP.
Therefore, not only their location but also their inference request rate is considered when clustering. 
If the number of inference requests exceeds available capacity, the local aggregator forwards the request to the global aggregator.

For the experiments, we assume that the latency for sending requests to the global server/cloud is between 50 and 100 ms. 
To make reasonable assumptions on latency, we measured HTTP response times (round-trip) to different cloud servers from other devices and networks. 
The latency cost to the local/edge servers is much lower and estimated between 8 and 10 ms.

\subsubsection{Inference serving latency}
\figurename~\ref{fig:answertimes} depicts the time in seconds it took until a client received a response to an inference request.
This figure shows that inference serving latency using the vanilla/non-hierarchical architecture is much higher than the other methods.
The higher response times are because it takes much more time to send the requests to the cloud server than to edge servers.
In these tests, we assumed that the edge servers had comparable processing power for the inference requests.
\vspace{-0.5cm}
\begin{figure}[htbp]
\centering
\includegraphics[width=0.8\linewidth, trim=0 0 50 80, clip]{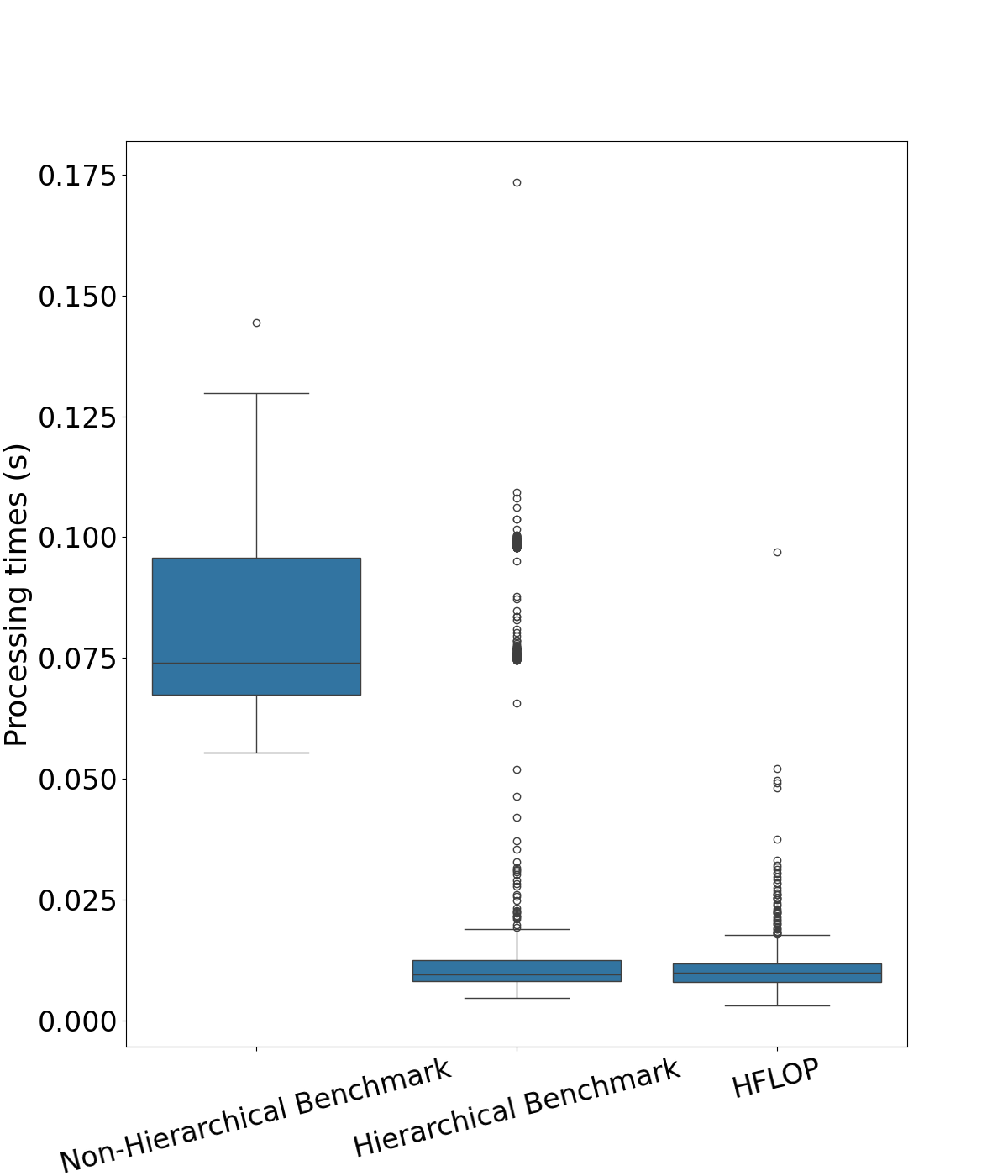}
\caption{Response times of inference requests.}
\label{fig:answertimes}
\end{figure}

The average response times were for the non-hierarchical benchmark 79.07 ms $\pm$ 15.94, for the hierarchical benchmark 17.72 ms $\pm$ 24.26, and for HFLOP 9.89 ms $\pm$ 4.63.

\subsubsection{End-to-end latency across different computing capacities}
This section compares the end-to-end latency when assuming different computing capacities on edge and cloud servers.
In the previous section, the assumption was made that the servers placed on edge nodes are as powerful as cloud servers.
Since this is often not the case, we compare the methods, considering a theoretical speedup of up to 95\%.

\begin{figure}[htbp]
    \centering
    \begin{subfigure}[b]{0.23\textwidth}
        \centering
        \includegraphics[width=\linewidth]{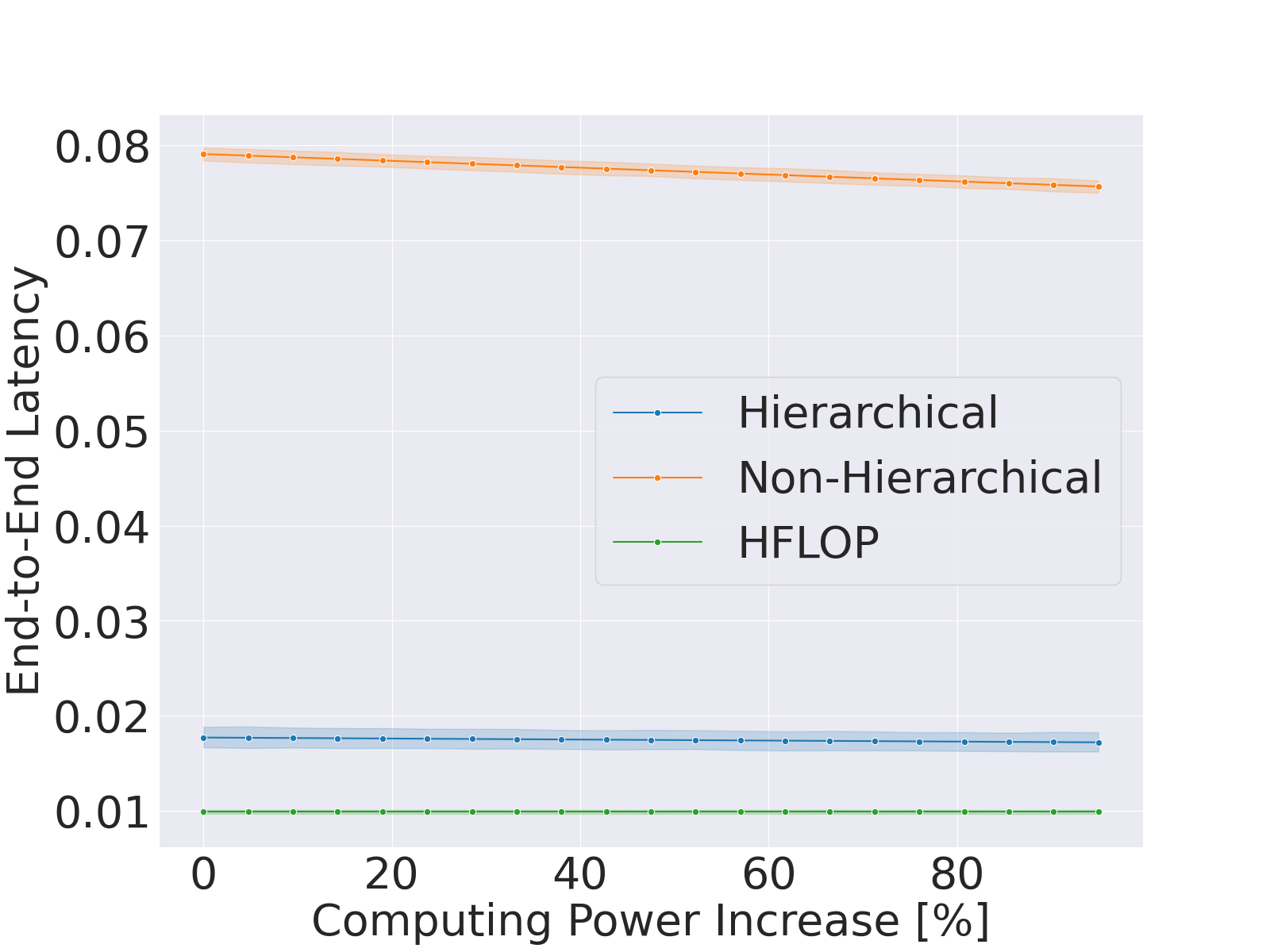}
        \caption{End-to-end latency with speedup and inference request rates $\lambda_i, \;\forall \; i\in[1,n]$.}
        \label{fig:speedups_1}
    \end{subfigure}
    \centering
    \begin{subfigure}[b]{0.24\textwidth}
        \centering
    \includegraphics[width=\linewidth]{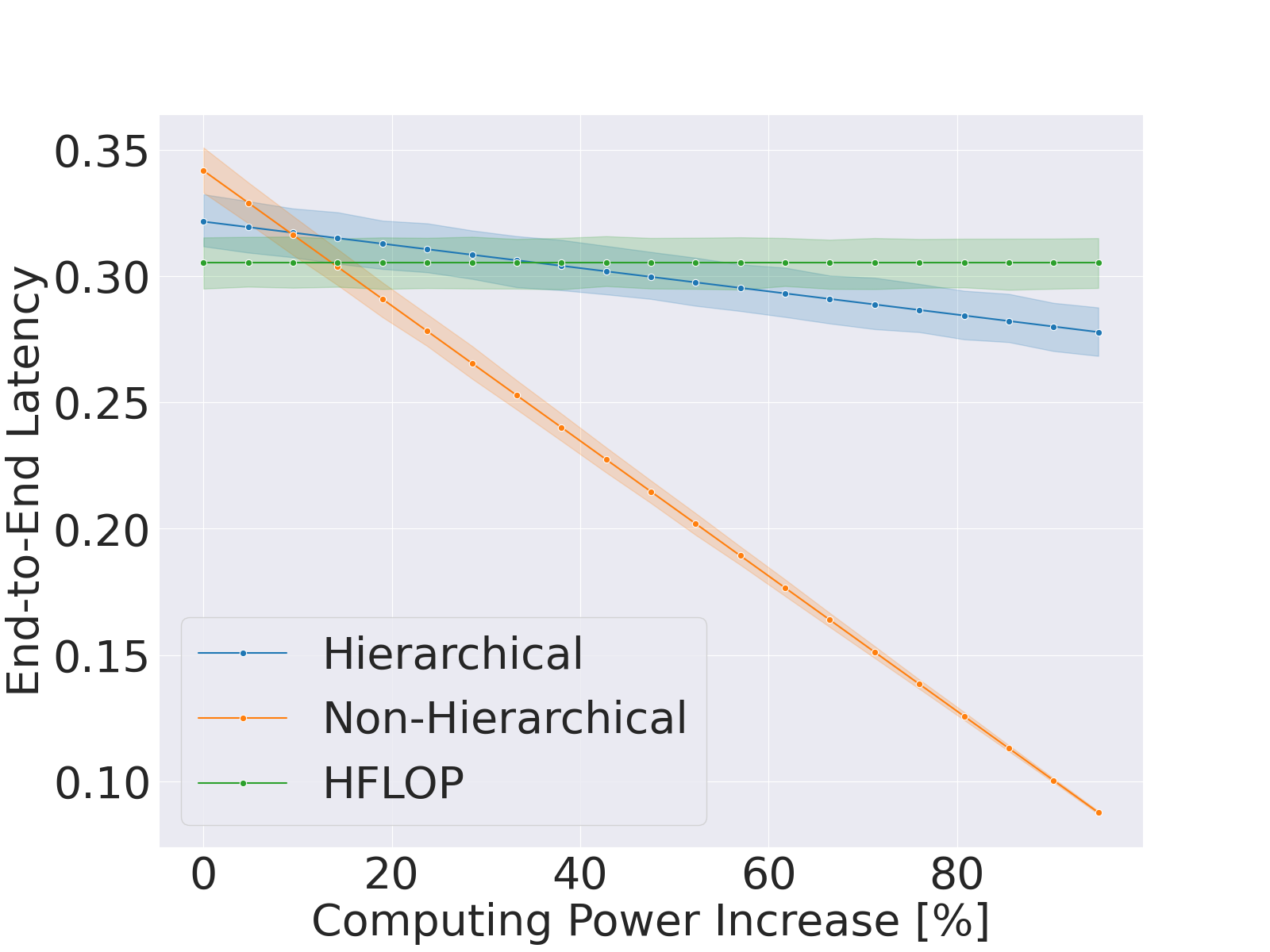}
        \caption{End-to-end latency with speedup and inference request rates $\lambda_i\times 10, \;\forall \; i \in[1,n]$.}
        \label{fig:speedups_10}
    \end{subfigure}
    \caption{End-to-end latency with different speedups and inference request rates.}
    \label{fig:speedups_overview}
\end{figure}
In \figurename~\ref{fig:speedups_overview},  one can see the end-to-end latency of the three methods.
The non-hierarchical benchmark is depicted in orange, the hierarchical benchmark is visualized in blue, and HFLOP is visualized in green.
Since the inference time is much lower than network-induced latency, there is almost no difference when increasing the speedup in \figurename~\ref{fig:speedups_1}.
When the inference request rate is increased 10 times (\figurename~\ref{fig:speedups_10}), the non-hierarchical method has a lower average latency than the other methods when the speedup is above 14.25.

\subsection{Cost savings}
We then turn our attention to communication costs. 
We simulate a scenario with $n = 500$ devices and increasing edge node densities with the following setup: For each device, there is a single edge node accessible at zero cost (e.g., a host in the same local area network), while the rest are accessible at unit cost. 
All edge hosts communicate with the global server at unit cost, and we force all clients to participate in training. Inference workloads and capacities are drawn uniformly at random. We compare the following three mechanisms: (i) vanilla FL, (ii) HFLOP, and (iii) the uncapacitated variant of HFLOP, where each edge host is assumed to have infinite request processing capacity. 
The ML model under training is the one applied in our use case, whose size in serialized format is 594\,KB. This is the payload uploaded/downloaded by an entity that submits a model update and receives a new model version. We configure HFL so that one global aggregation round takes place every two local ones; this is a rather conservative approach from a cost perspective, as it leads to a larger number of costly global aggregations. We compare the three approaches in terms of their communication cost \emph{until convergence}, expressed as \emph{the volume of traffic exchanged over metered links}, i.e., excluding the traffic exchanged over zero-cost connections. In the experiments presented in \figurename~\ref{fig:global}, convergence is ensured after approximately 100 aggregation rounds in all scenarios, which is what we also assume for this experiment. (If we assume convergence after 20 rounds instead, as discussed in Section~\ref{sec:continuous-fl}, the results in absolute traffic volume terms we report below are scaled down accordingly.) Given our configuration, this corresponds to 50 global aggregation rounds for the hierarchical FL experiments. The uncapacitated HFLOP variant serves as the lower bound in terms of communication cost: since inference workloads of devices and the inference processing capacity of individual edge nodes are ignored, the derived solution is the best one could expect from a cost perspective.

\begin{figure}
    \centering
    \includegraphics[width=\columnwidth]{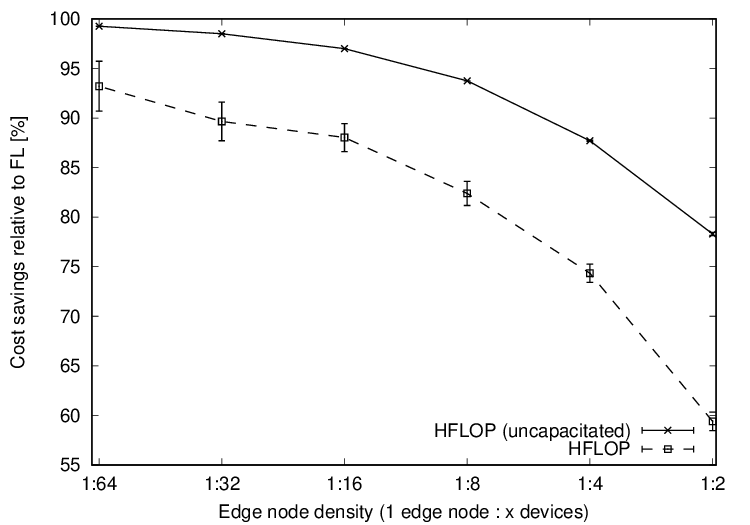}
    \caption{Cost savings relative to standard FL for increasing edge node densities and a fixed number of $n = 200$ devices. Mean values are reported with 95\% confidence intervals.}
    \label{fig:cost-benchmark}
\end{figure}

\figurename~\ref{fig:cost-benchmark} reports the communication cost reduction (\%) the two HFL variants achieve against standard FL. In our setup, these savings are more drastic when edge host density is low: devices always have an edge host reachable at zero cost, which may be used if capacity constraints (if any) allow it and costly aggregator-cloud links are only a few. In practice, though, this would necessitate a few very high-capacity \emph{and} cheaply reachable edge nodes. 
In more general settings (i.e., where devices and edge nodes have diverse communication costs and not unit ones), the cost gap between the two HFL variants is much more narrow.
Also, this cost gap gets narrower the more the overall available capacity; the capacity configurations we selected for this experiment favor the uncapacitated version of HFLOP. 
Finally, in absolute traffic volume terms and for the clustered topology we experimented with in our use case (4 edge nodes, 20 devices), approximately 2.37\,GB, 0.53\,GB, and 0.24\,GB was transmitted over metered links until GRU model training convergence using FL, HFLOP, and its uncapacitated version, respectively.

\section{Discussion}
\noindent\textbf{Variants of the orchestration problem.}
HFLOP can be extended in various ways to capture different operational requirements. Straightforward extensions include enforcing privacy-related constraints, where a device is allowed to associate only with edge nodes that it trusts, and considering device reliability in assignment decisions as long as such per-device information exists. These can be implemented with modified or additional HFLOP constraints. More importantly, current hierarchical federated learning designs often consider the effects of differences in client data distributions on learning performance, adding this dimension in their configuration decisions~\cite{Deng24}. Solving HFLOP while ensuring balanced data distributions among device clusters is an interesting and challenging problem.

\noindent\textbf{Alternatives for inference serving.}
In our problem settings, we assume that a device that is busy training can offload all its inference workload to its associated edge node. Other inference-serving schemes, however, can be considered. For instance, trading inference quality for fast local execution, a device whose GPU is busy with an FL task can use a lower-complexity ML model for the same inference task suitable for CPU execution, such as a quantized version of the global model. Furthermore, more intelligent request routing strategies could be implemented at the device and the edge host (aggregator) end.

\noindent\textbf{Dealing with environment dynamics.}
In practical FL settings, it is reasonable to expect that the device population fluctuates and that the conditions at edge nodes may change, e.g., in the presence of other application workloads that may affect the request processing capacity of edge nodes. Such changes may imply that a solution to HFLOP is not optimal anymore, and the question of what conditions should trigger its re-execution remains open. Adapting an HFL configuration in response to such changes requires considering reconfiguration costs and potentially applying heuristic schemes to deal with the complexity of HFLOP. This is the subject of our ongoing work on \emph{adaptive FL orchestration}.

\section{Conclusion}
\label{sec:conclusion}

This work addressed the challenge of inference serving during training in a federated learning environment. This challenge is particularly evident in continuous learning settings, where models must be retrained while serving inference requests. To this end, we introduced the inference-aware Hierarchical Federated Learning Orchestration Problem (HFLOP), which clusters federated learning clients considering their individual inference request workload and network costs, ensuring that inference requests are served as close to the clients as possible, lowering communication costs and response times. 
Our systematic experiments show that our orchestration scheme decreases the communication cost without affecting the continuous training of the federated learning clients.
Our experiments also demonstrated significantly reduced end-to-end inference latency, achieving better results by processing inference requests on nearby edge servers. These performance benefits were sustained even for inference processing workloads that exceeded edge infrastructure capacity and in the presence of highly asymmetric compute resource capabilities between edge and cloud hosts.

\section*{Acknowledgment} This work has been supported by the European Union's Horizon Europe research and innovation program under grant agreements No. 101079214 (AIoTwin) and No. 101135576 (INTEND).

% Generated by IEEEtran.bst, version: 1.14 (2015/08/26)

\end{document}